%% file: main.tex
\documentclass[a4paper,11pt]{article}
\usepackage{pos}
\usepackage{xspace}
\newcommand{\pysecdec}{py{\textsc{SecDec}}\xspace}
\usepackage{subcaption}
\usepackage{tikz}
\usepackage[compat=1.1.0]{tikz-feynman}
\usepackage{tikz-3dplot}
\usepackage{axodraw2}
\usetikzlibrary{calc}
\usetikzlibrary{arrows.meta}
\usetikzlibrary{decorations.markings}
\usetikzlibrary{decorations.pathmorphing}
\usetikzlibrary{shapes.geometric}
\pgfdeclarelayer{nodelayer}
\pgfdeclarelayer{edgelayer}
\pgfsetlayers{main,edgelayer,nodelayer}
\tikzstyle{none}=[]
\input{all.tikzstyles}

\title{Accelerating Feynman Integral Evaluation by Avoiding Contour Deformation}

\author[a]{Stephen P. Jones}
\author*[a]{Anton Olsson}
\author[b]{Thomas Stone}

\affiliation[a]{Institute for Particle Physics Phenomenology, \\ Durham University, Durham DH1 3LE, UK}
\affiliation[b]{Physics Department, Technical University of Munich,\\
James-Franck-Straße 1, 85748 Garching, Germany}

\emailAdd{stephen.jones@durham.ac.uk}
\emailAdd{anton.olsson@durham.ac.uk}
\emailAdd{thomas.stone@tum.de}

\abstract{We describe our method for rewriting dimensionally regulated Feynman parameter integrals in the Minkowski regime as a sum of real, positive integrands multiplied by complex prefactors. This representation eliminates the need for a contour deformation, which is one of the main computational bottlenecks in numerical integration. We demonstrate clearly how the method works on two examples, and benchmark the performance against contour deformation as implemented in \pysecdec, where we observe performance gains of up to several orders of magnitude. We describe an improvement in the resolution procedure using the Generic Cylindrical Algebraic Decomposition algorithm, which generalises our method to any Feynman integral, including those with massive propagators.}

\FullConference{17th International Symposium on Radiative Corrections: Applications of Quantum Field Theory to Phenomenology (RADCOR2025)\\
5-10 October 2025\\
Puri, India\\}

\begin{document}
\maketitle

\section{Introduction}
Higher-order radiative corrections to scattering processes involve the computation of multi-loop amplitudes, which require the evaluation of multi-loop Feynman integrals. Feynman integrals with enough loops and kinematic scales are analytically intractable with current methods, but are still required for phenomenological applications. In these cases, numerical techniques provide a viable and often essential alternative.

Numerical evaluations of Feynman integrals for physical scattering kinematics necessitate a prescription for resolving integrable singularities, that are encountered inside the integration domain. Traditionally, this is achieved by applying a contour deformation, which avoids the singularities by shifting the integration contour into the complex plane~\cite{Soper:1998ye,Soper:1999xk,Binoth:2005ff,Nagy:2006xy,Anastasiou:2006hc,Anastasiou:2007qb,Lazopoulos:2007ix,Lazopoulos:2007bv,Anastasiou:2008rm,Gong:2008ww,Becker:2010ng,Mizera:2021icv,Hannesdottir:2022bmo,Borinsky:2023jdv}. Contour deformation has been a standard strategy for a long time, as it is a widely applicable method that works on most integrals, and is easily automated~\cite{Borowka:2014aaa,Borowka:2015mxa,Borowka:2017idc,Borowka:2018goh,Jahn:2020tpj,Heinrich:2021dbf,Heinrich:2023til,Smirnov:2013eza,Smirnov:2015mct,Smirnov:2021rhf,Borinsky:2023jdv}. The downside is that for complicated integrals with many propagators, it adds a significant computational load and strongly limits the the class of integrals that can be evaluated efficiently. 

In these proceedings we describe an alternative method for evaluating Feynman integrals for physical scattering kinematics, that does not require a contour deformation~\cite{Jones:2024gmw,Jones:2025jzc}. This method was first presented in our main paper, Ref.~\cite{Jones:2025jzc}, and the goal of these proceedings is to provide a concise, pedagogical overview of the techniques discussed therein, and to describe some recent improvements. 

\section{Feynman integrals}
\label{sec:preliminaries}
A dimensionally regularised Feynman integral with $L$ loops and $D=4-2\epsilon$ dimensions can be written in terms of Feynman parameters as, 
\begin{equation}
J(\mathbf{s})=\frac{\left(-1\right)^{\nu}\Gamma\left(\nu-L D/2\right)}{\prod_{i=1}^{N}\Gamma\left(\nu_{i}\right)}\lim_{\delta\to0^+}\!\int_{\mathbb{R}_{\geq0}^{N}}\prod\limits_{i=1}^{N}\mathrm{d}x_{i} x_{i}^{\nu_{i}-1}\frac{\mathcal{U}\!\left(\mathbf{x}\right)^{\nu-(L+1)D/2}}{\left(\mathcal{F}\!\left(\mathbf{x};\mathbf{s}\right)-i\delta\right)^{\nu-LD/2}}\delta(1-\alpha(\mathbf{x})),
\label{eq:fp2}
\end{equation}
where $\mathcal{U}(\mathbf{x})$ and $\mathcal{F}(\mathbf{x};\mathbf{s})$ are respectively degree $L$ and $L+1$ homogeneous polynomials in the Feynman parameters, $\mathbf{x}=(x_1,\ldots,x_N)$. The $\delta>0$ imposes the causal prescription for Feynman propagators, the $\nu_i$ are powers of the propagators and $\nu = \sum_{i=1}^N \nu_i$. The integral is bounded by $\delta(1-\alpha(\mathbf{x}))$, where $\alpha(\mathbf{x})$ is a hyperplane in the positive domain, see Refs.~\cite{Cheng:1987ga,Panzer:2015ida,Weinzierl:2022eaz}. The integral depends on external kinematic invariants, $s_j$, and internal masses, $m_i$, via $\mathbf{s}=(s_1,\ldots,s_M,m^2_1,...,m^2_N)$. It is useful to introduce two kinematic regimes, depending on the sign of the $\mathcal{F}(\mathbf{x};\mathbf{s})$ polynomial. 
If the kinematic configuration is such that $\mathcal{F}(\mathbf{x};\mathbf{s}) > 0$ everywhere inside the integration domain, we call such a configuration Euclidean\footnote{If $\mathcal{F}(\mathbf{x};\mathbf{s}) < 0$ everywhere this can trivially be brought to Euclidean form by factoring $(-1 - i\delta)$ out of the $\mathcal{F}(\mathbf{x};\mathbf{s})$ polynomial.}. 
If instead the sign of the kinematic invariants are such that the $\mathcal{F}(\mathbf{x};\mathbf{s})$ polynomial can be both positive and negative, in different regions of the parameter space, we call this the Minkowski regime.

A Feynman integral is characterised by its singularities, which are captured by the Landau equations. In parameter space, they can be written in the form of the following two conditions~\cite{Eden:1966dnq}
\begin{equation}
    \mathcal{F}(\mathbf{x};\mathbf{s}) = 0, \qquad x_k \frac{\partial \mathcal{F}(\mathbf{x};\mathbf{s})}{\partial x_k} = 0, \quad \text{for all } k \in \{1,\dots,N\}.
    \label{eq:landau}
\end{equation}
These are necessary but not sufficient conditions for a Feynman integral to have a singularity. The second condition implies two possibilities; either $\mathcal{F}(\mathbf{x};\mathbf{s})$ vanishes at the integration boundary, or at the same time as its derivatives vanish. In the former case, the singularities can be algorithmically extracted using sector decomposition \cite{Binoth:2000ps,Binoth:2003ak,Heinrich:2008si,Kaneko:2009qx,Schlenk:2016epj,Heinrich:2021dbf}. In the latter case, it has been shown recently how also such singularities can be extracted with sector decomposition by first splitting the Newton polytope on the singular locus \cite{Gardi:2024axt}.

\subsection{Contour deformation}
In these proceedings, we are interested in the situation where only the first condition of Eq.~\eqref{eq:landau} is fulfilled, i.e. $\mathcal{F}(\mathbf{x};\mathbf{s}) = 0$, but away from the boundary of integration and while at least some partial derivatives do not vanish. In this case, the singularity is integrable and can, for example, be resolved by implementing a contour deformation. This means that we shift each integration variable away from the real line by a small imaginary part,
\begin{equation}
    x_k \to z_k = x_k - i\tau_k,
\label{eq:deformation-shift}
\end{equation}
where the $\tau_k$ must be chosen in accordance with the causal $i\delta$ prescription. The advantage of this strategy is that it is very algorithmic and provides a general, automatable method for defining Feynman integrals in the Minkowski regime. Various versions of contour deformation has been implemented in many public numerical codes \cite{Borowka:2014aaa,Borowka:2015mxa,Borowka:2017idc,Borowka:2018goh,Jahn:2020tpj,Heinrich:2021dbf,Heinrich:2023til,Smirnov:2013eza,Smirnov:2015mct,Smirnov:2021rhf,Borinsky:2023jdv}. 

There are, however, several drawbacks of using a contour deformation. The main issue is that after the shift in Eq.~\eqref{eq:deformation-shift}, the integrand receives both positive and negative contributions from different regions of the parameter space, which leads to cancellations and thus a loss of numerical precision. These cancellations are a problem for most phase-space points, but tend to be more severe in extreme kinematic configurations, such as in high-energy or small-mass limits. 

\begin{figure}[h!]
    \centering
    \includegraphics[width=\linewidth]{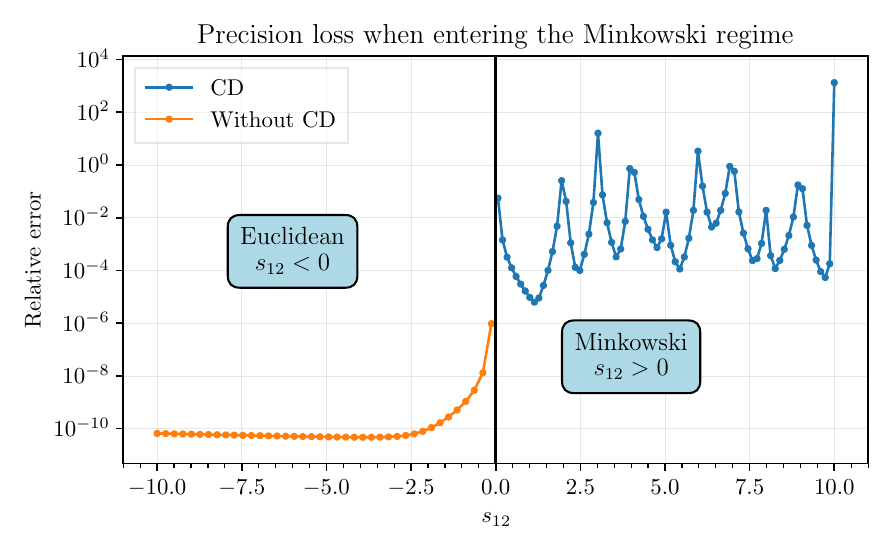}
    \caption{Precision loss due to contour deformation, when integrating the one-loop pentagon integral for physical kinematic configurations. The orange points correspond to Euclidean configurations and can be evaluated without any resolution procedure. The other invariants are fixed at $(s_{23}, s_{34}, s_{45}, s_{51})~=~(-4, -2, -6, -3)$, and each point is evaluated with \pysecdec using $10^5$ QMC samples. In addition to the reduced precision, the blue points (using contour deformation) are $\approx 2.67\times$ slower to evaluate than the orange points, for a fixed number of QMC points.}
    \label{fig:cd_precision_loss}
\end{figure}

We demonstrate this in Figure~\ref{fig:cd_precision_loss}, where we have used \pysecdec~\cite{Borowka:2014aaa,Borowka:2015mxa,Borowka:2017idc,Borowka:2018goh,Jahn:2020tpj,Heinrich:2021dbf,Heinrich:2023til} to evaluate a massless one-loop pentagon integral for different values of the centre-of-mass energy $s_{12} \in (-10, 10)$, with a fixed number of Quasi-Monte Carlo (QMC) points. When $s_{12} > 0$, the integral enters the Minkowski regime and requires a contour deformation. We observe a loss of precision of five digits or more, for most phase-space points. Moreover, as the contour-deformed integrand is much more complicated\footnote{After the shift in Eq.~\eqref{eq:deformation-shift}, the integration variables are transformed back to the real line which introduces an $(N-1) \times (N-1)$ Jacobian determinant, which often comprises the bulk of the integrand after deformation.}, each integrand evaluation is slower, in this case by a factor of $2.67$. For the same reason, the integration libraries for contour deformed integrands are typically much larger, which can cause practical problems during generation and compilation.

Furthermore, the exact choice of deformation parameters in Eq.~\eqref{eq:deformation-shift} can significantly impact numerical performance, and selecting optimal deformation parameters is a highly non-trivial task~\cite{Winterhalder:2021ngy,Jones:2026tkf}. In special cases, no value of the deformation parameters would produce a valid contour, due to the presence of Landau singularities within the integration domain~\cite{Gardi:2024axt}.

In summary, contour deformation is slow, arbitrary and known to fail. In these proceedings we present an alternative strategy for resolving integrable singularities of Feynman integrals, by decomposing the integral into a sum of real, non-negative contributions multiplied by complex prefactors. 

\section{Method}
\label{sec:method}
The idea of the new method is to split the integration domain into two or more regions based on the sign of the $\mathcal{F}(\mathbf{x};\mathbf{s})$ polynomial. For brevity, we use the notation $\mathrm{d}\mathbf{x} = \prod_{i=1}^{N}\mathrm{d}x_{i}$ and introduce
\begin{equation}
    I(\mathbf{s}; \delta) = \int_{\mathbb{R}_{\geq0}^{N}}\prod\limits_{i=1}^{N}\mathrm{d}x_{i} x_{i}^{\nu_{i}-1}\frac{\mathcal{U}\!\left(\mathbf{x}\right)^{\nu-(L+1)D/2}}{\left(\mathcal{F}\!\left(\mathbf{x};\mathbf{s}\right)-i\delta\right)^{\nu-LD/2}}\delta(1-\alpha(\mathbf{x})) = \int_{\mathbb{R}_{\geq0}^{N}} \mathrm{d}\mathbf{x} \, \mathcal{I}(\mathbf{x};\mathbf{s}),
\end{equation}
such that $J(\mathbf{s}) = \frac{\left(-1\right)^{\nu}\Gamma\left(\nu-L D/2\right)}{\prod_{i=1}^{N}\Gamma\left(\nu_{i}\right)} \lim_{\delta\to0^+} I(\mathbf{s};\delta)$. We now decompose the integration domain into a piece where $\mathcal{F}(\mathbf{x};\mathbf{s}) > 0$ and one where $\mathcal{F}(\mathbf{x};\mathbf{s}) < 0$,
\begin{equation}
\label{eq:initial_decomp}
    I(\mathbf{s};\delta) = \int_{\mathbb{R}_{\geq0}^{N}} \mathrm{d}\mathbf{x} \, \mathcal{I}(\mathbf{x};\mathbf{s}) = \int_{\mathbb{R}_{\geq0}^{N}} \mathrm{d}\mathbf{x} \, \mathcal{I}(\mathbf{x};\mathbf{s}) \left[\theta(\mathcal{F}\!\left(\mathbf{x};\mathbf{s}\right)) + \theta(-\mathcal{F}\!\left(\mathbf{x};\mathbf{s}\right)) \right],
\end{equation}
where $\theta(x)$ is the Heaviside step function. $\mathcal{F}(\mathbf{x};\mathbf{s}) > 0$ and $\mathcal{F}(\mathbf{x};\mathbf{s}) < 0$ correspond to regions in parameter space 
and can be represented by a set of inequalities involving the integration variables and the kinematic invariants. If these inequalities are taken as the new boundaries of integration, we can write the right-hand side of Eq.~\eqref{eq:initial_decomp} as,
\begin{equation}
    \int_{\mathbb{R}_{\geq0}^{N}} \mathrm{d}\mathbf{x} \, \mathcal{I}(\mathbf{x};\mathbf{s}) \left[\theta(\mathcal{F}\!\left(\mathbf{x};\mathbf{s}\right)) + \theta(-\mathcal{F}\!\left(\mathbf{x};\mathbf{s}\right)) \right] = \int_{S^+} \mathrm{d}\mathbf{x} \, \mathcal{I}(\mathbf{x};\mathbf{s}) + \int_{S^-} \mathrm{d}\mathbf{x} \, \mathcal{I}(\mathbf{x};\mathbf{s}),
\end{equation}
where $S^+$ ($S^-$) is the set of inequalities that result from reducing $\mathcal{F}(\mathbf{x};\mathbf{s}) > 0$ ($\mathcal{F}(\mathbf{x};\mathbf{s}) < 0$). In Ref.~\cite{Jones:2025jzc}, we focused on so-called Univariate Bisectable (UB) integrals, where $S^{+}$ is a constraint on just one integration variable $x_i$. In this case, $S^+$ is either of (and $S^-$ is the other),
\begin{equation}
    \{0<x_i<f\left(\mathbf{x}_{\neq i}\right)\} \quad \mathrm{or} \quad \{f\left(\mathbf{x}_{\neq i}\right)<x_i\},
\end{equation}
where $\mathbf{x}_{\neq i}$ denotes a dependence on all integration variables except $x_i$, and $f\left(\mathbf{x}_{\neq i}\right)$ is a rational (algebraic) function if $\mathcal{F}(\mathbf{x};\mathbf{s})$ is linear (quadratic) in $x_i$.  
In each case above, the original boundary of integration is restored by applying the variable transformations
\begin{equation}
\label{eq:transformations}
    x_i'=\frac{x_ix_j}{f\left(\mathbf{x}_{\neq i}\right)-x_i} \quad \mathrm{or} \quad x_i'=x_i-f\left(\mathbf{x}_{\neq i}\right),
\end{equation}
which remap the $x_i = f\left(\mathbf{x}_{\neq i}\right)$ hypersurface to $\infty$ and $0$ respectively. The result is a decomposition,
\begin{equation}
    I(\mathbf{s};\delta) = \int_{\mathbb{R}_{\geq0}^{N}} \mathrm{d}\mathbf{x} \, \mathcal{I}^+(\mathbf{x};\mathbf{s}) + \int_{\mathbb{R}_{\geq0}^{N}} \mathrm{d}\mathbf{x} \, \mathcal{I}^-(\mathbf{x};\mathbf{s}),
    \label{decompositionequation}
\end{equation}
where $\mathcal{I}^{\pm}(\mathbf{x};\mathbf{s})$ are the integrands resulting from applying the transformations in Eq.~\eqref{eq:transformations} and including the appropriate Jacobians. The integrable singularity at $\mathcal{F}(\mathbf{x};\mathbf{s}) = 0$ has in each case been mapped to the boundary, and can now be subtracted algorithmically using sector decomposition. Inside the integration domain, each integrand is of uniform sign, $\mathcal{I}^{+}(\mathbf{x};\mathbf{s}) > 0$ and $\mathcal{I}^{-}(\mathbf{x};\mathbf{s})~<~0$. We can factor out a minus sign from the denominator polynomial in $\mathcal{I}^{-}(\mathbf{x};\mathbf{s})$, to obtain exclusively non-negative integrands\footnote{In Ref.~\cite{Jones:2025jzc}, the minus sign was already factored out of the integral in the definition of $\mathcal{I}^{-}(\mathbf{x};\mathbf{s})$.}. The result for the full integral, $J(\mathbf{s})$, is then
\begin{equation}
    J(\mathbf{s})=\sum_{n_+=\ \!\!1}^{N_+}J^{+,n_+}(\mathbf{s})+\lim_{\delta\to0^+}\left(-1-i\delta\right)^{-\left(\nu-LD/2\right)}\sum_{n_-=\ \!\!1}^{N_-}J^{-,n_-}(\mathbf{s}),
    \label{eq:decomp}
\end{equation}
where the analytic continuation is now fully determined by the prefactor $\left(-1-i\delta\right)^{-\left(\nu-LD/2\right)}$ and the $J^{+,n_+}(\mathbf{s})$, $J^{-,n_-}(\mathbf{s})$ are integrals over real, non-negative integrands. The sums over $n_+$ and $n_-$ indicate the generalisation to more complicated examples, where the positive and negative regions may need to be further decomposed. 

The main challenge of the method described above is in reducing $\mathcal{F}(\mathbf{x};\mathbf{s}) > 0$ and $\mathcal{F}(\mathbf{x};\mathbf{s}) < 0$ to a set of inequalities in $\mathbf{x}$. In the literature, there are algorithms that reduce systems of polynomial inequalities which are known to terminate. One such algorithm is the Generic Cylindrical Algebraic Decomposition (GCAD)~\cite{strictineqs}, a variant of CAD~\cite{gcad} relevant for our purposes, which is available in \texttt{Mathematica}.
In Section~\ref{sec:examples}, we use GCAD to write Feynman integrals in the form of Eq.~\eqref{eq:decomp}, and demonstrate the benefit of this representation for numerical integration.

\section{Examples}
\label{sec:examples}
In this section we apply the method described in Section~\ref{sec:method} on two examples and demonstrate the benefit of the representation in Eq.~\eqref{eq:decomp} for numerical integration. We first carefully describe how the method works in practice for a massless two-loop non-planar box. We follow by demonstrating how using GCAD has yielded an improvement in the resolution procedure for the all-massive triangle. We refer to our main paper, Ref.~\cite{Jones:2025jzc}, for a more extensive list of examples, which includes massless and massive integrals up to three loops.
\begin{figure}[t]
    \centering
    \begin{subfigure}[t]{0.49\textwidth} 
    \centering
    \begin{tikzpicture}[baseline=13ex,scale=1.0]
        \coordinate (x1) at (1, 3) ;
        \coordinate (x2) at (1, 1) ;
        \coordinate (x3) at (2, 2) ;
        \coordinate (x5) at (3, 1) ;
        \coordinate (x4) at (3, 3) ;
        \node (p1) at (0, 3) {$p_1$};
        \node (p2) at (0, 1) {$p_2$};
        \node (p3) at (1.33, 1.33) {$p_3$};
        \node (p4) at (4, 3) {$p_4$};
        \draw[color=blue] (x1) -- (p1);
        \draw[color=blue] (x2) -- (p2);
        \draw[color=blue] (x3) -- (p3);
        \draw[color=blue] (x4) -- (p4);
        \draw[ultra thick,color=Black] (x1) -- (x2) node [midway,,xshift=-7pt,yshift=0,color=Black] {$x_1$};
        \draw[ultra thick,color=Black] (x1) -- (x4) node [midway,,xshift=0pt,yshift=7pt,color=Black] {$x_3$};
        \draw[ultra thick,color=Black] (x2) -- (x5) node [midway,,xshift=0pt,yshift=-7pt,color=Black] {$x_6$};
        \draw[ultra thick,color=Black] (x3) -- (x1) node [midway,,xshift=+9pt,yshift=+1.6pt,color=Black] {$x_2$};
        \draw[ultra thick,color=Black] (x3) -- (x5) node [midway,,xshift=-6pt,yshift=-5pt,color=Black] {$x_5$};
        \draw[ultra thick,color=Black] (x4) -- (x5) node [midway,,xshift=7.75pt,yshift=0pt,color=Black] {$x_4$};
        \draw[fill,thick,color=Blue] (x1) circle (1pt);
        \draw[fill,thick,color=Blue] (x2) circle (1pt);
        \draw[fill,thick,color=Blue] (x3) circle (1pt);
        \draw[fill,thick,color=Blue] (x4) circle (1pt);
        \draw[fill,thick,color=Blue] (x5) circle (1pt);
    \end{tikzpicture}
    \label{massless_BNP6}
    \end{subfigure}
    \hspace{-2.3cm}
    \begin{subfigure}[t]{0.49\textwidth}
    \centering
    \begin{tikzpicture}[baseline=13ex,scale=1.0]
            \coordinate (x1) at (0.5, 1) ;
            \coordinate (x2) at (2.5, 1) ;
            \coordinate (x3) at (1.5,2.5) ;
            \node (p1) at (-0.1, 0.6) {};
            \node (p2) at (3.1, 0.6) {};
            \node (p3) at (1.5,3.5) {};
            \node (p4) at (1.1,3) {$p^2$};
            \draw[color=blue] (x1) -- (p1);
            \draw[ultra thick,color=ForestGreen] (x3) -- (p3);
            \draw[color=blue] (x2) -- (p2);
            \draw[ultra thick,color=Purple] (x1) -- (x2) node [midway,yshift=+6pt,color=Black] {$m$};
            \draw[ultra thick,color=Purple] (x2) -- (x3) node [midway,xshift=+8pt,color=Black] {$m$};
            \draw[ultra thick,color=Purple] (x3) -- (x1) node [midway,xshift=-9pt,color=Black] {$m$};
            \draw[fill,thick,color=Blue] (x1) circle (1pt);
            \draw[fill,thick,color=Blue] (x2) circle (1pt);
            \draw[fill,thick,color=Blue] (x3) circle (1pt);
        \end{tikzpicture}
        \label{fullmasstri}
    \end{subfigure}
    \caption{Feynman diagrams for a two-loop non-planar box (BNP6) and the all-massive one-loop triangle.}
    \label{fig:examples}
\end{figure}
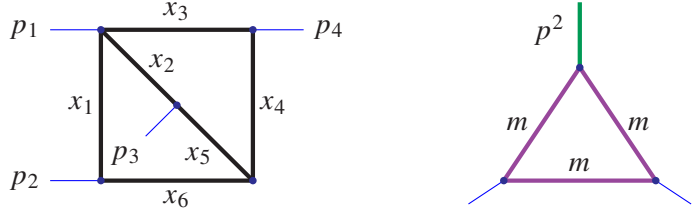

\subsection{Two-loop non-planar box}
\begin{figure}[h!]
\centering
    \includegraphics[width=0.9\textwidth]{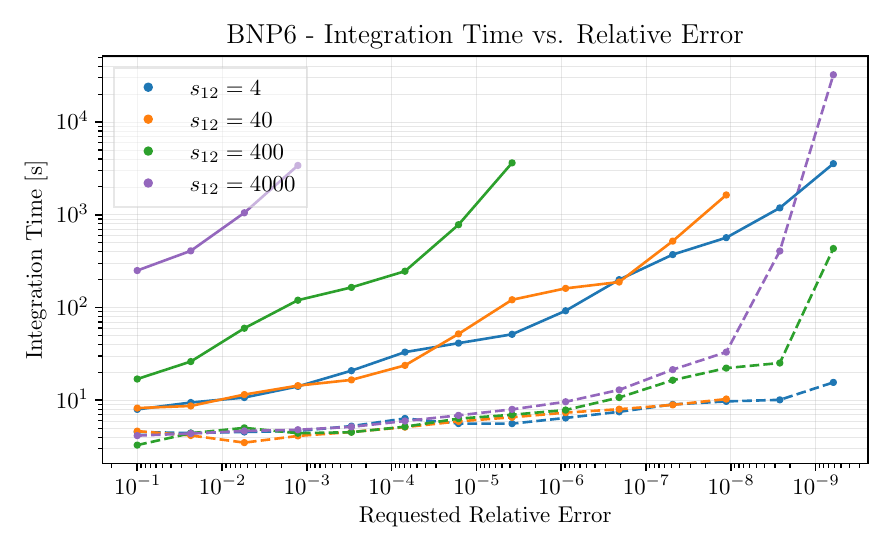}
    \caption{Timings with (solid lines) and without (dashed lines) contour deformation for the two-loop non-planar box with six propagators, expanded up to the finite order. Evaluated for different values of $s_{12}$ with $s_{23}=-1$ fixed.}
    \label{fig:BNP6_timings}
\end{figure}

We start by considering a non-planar two-loop box with six propagators (BNP6, see Figure~\ref{fig:examples}) which can be parameterised by the Mandelstam invariants $s_{12}=(p_1+p_2)^2$ and $s_{23} = (p_2+p_3)^2$ after applying the momentum conservation rule $s_{12}+s_{23}+s_{13}=0$ to eliminate $s_{13}$. 
The integral may be written as,
\begin{align}
    J_{\mathrm{BNP6}}(\mathbf{s})&=\Gamma\left(2+2\epsilon\right)\lim_{\delta\to0^+}I_{\mathrm{BNP6}}(\mathbf{s};\delta),\\I_{\mathrm{BNP6}}(\mathbf{s};\delta)&=\int_{\mathbb{R}^6_{\geq0}}\prod_{i=1}^6\mathrm{d}x_i\, \frac{\mathcal{U}(\mathbf{x})^{3\epsilon}}{\left(\mathcal{F}(\mathbf{x}; \mathbf{s}) - i\delta\right)^{2+2\epsilon}}\delta\left(1-\alpha(\mathbf{x})\right)
\label{BNP6-integral}
\end{align}
with the $\mathcal{U}$ and $\mathcal{F}$ polynomials,
\begin{align}
\mathcal{U}(\mathbf{x}) = \, &x_1x_2 + x_1x_3 + x_1x_4 + x_1x_5 + x_2x_3 + x_2x_4 + x_2x_6 + \nonumber \\ 
&x_3x_5 + x_3x_6 + x_4x_5 + x_4x_6 + x_5x_6,\\
\mathcal{F}(\mathbf{x}; \mathbf{s}) = &-s_{12} x_2 x_3 x_6 -s_{23} x_1 x_2 x_4+(s_{12}+s_{23}) x_1 x_3 x_5.
\end{align}
We restrict to the physical kinematic regime for massless $2\rightarrow2$ scattering, $\mathbf{s}_\mathrm{phys}=\{0<s_{12}<\infty,\ -s_{12}<s_{23}<0\}$. This is an example of a UB integral, and applying GCAD to reduce $\mathcal{F}(\mathbf{x}; \mathbf{s})~<~0$, yields a constraint on just one integration variable, $x_1$. The constraint and corresponding transformation to map the upper boundary to infinity are\footnote{The $x_6$ in the denominator of the transformation in Eq.~\eqref{eq:BNP6_neg} is included to preserve homogeneity of the integrand after the transformation.}
\begin{equation}
\label{eq:BNP6_neg}
    0 < x_1 < f(\mathbf{x}_{\neq1}) = \frac{s_{12}x_2x_3x_6}{\left(s_{12}+s_{23}\right)x_3x_5-s_{23}x_2x_4}, \quad \quad x_1\rightarrow \frac{x_1}{x_1+x_6} f(\mathbf{x}_{\neq1}).
\end{equation}
Applying the transformation on the right-hand side of Eq.~\eqref{eq:BNP6_neg} to the $\mathcal{U}(\mathbf{x})$ and $\mathcal{F}(\mathbf{x}; \mathbf{s})$ polynomials, and including the Jacobian of the transformation, yields the resolved negative contribution where the minus sign has already been factored out,
\begin{align}
    &\mathcal{I}_{\mathrm{BNP6}}^-=\left(s_{12}x_2x_3x_6^2\right)^{-1-2\epsilon}\left(x_1+x_6\right)^{-\epsilon}\left[(s_{12}+s_{23})x_3x_5-s_{23}x_2x_4\right]^{-1-3\epsilon}\Bigl[[(s_{12}+s_{23})x_3 x_5 -\nonumber\\& s_{23} x_2 x_4](x_1+x_6) [(x_3+x_4) (x_2+x_5)+\left(x_2+x_3+x_4+x_5\right) x_6] + s_{12} x_1 x_2 x_3 x_6\left(x_2+x_3+x_4+x_5\right)\Bigr]^{3\epsilon}. \nonumber
\end{align}
Since this is a UB integral, the constraint for the positive contribution is the inverse of the above, $x_1 > f(\mathbf{x}_{\neq1})$, and $\mathcal{I}_{\mathrm{BNP6}}^+$ is obtained by applying the transformation $x_1 \rightarrow x_1 + f(\mathbf{x}_{\neq1})$, shifting the lower boundary to $0$. Combining the positive and negative contributions we can write $I_{\mathrm{BNP6}}(\mathbf{s};\delta)$ on the form of Eq.~\eqref{eq:decomp} as,
\begin{align}
    I_{\text{BNP6}}(\mathbf{s};\delta)=I^{+}_{\text{BNP6}}(\mathbf{s})+\left(-1-i\delta\right)^{-2-2\epsilon}I^{-}_{\text{BNP6}}(\mathbf{s}).
    \label{decompBNP6}
\end{align}
In Figure~\ref{fig:BNP6_timings}, we compare the integration time to obtain a certain requested precision, between the contour deformed BNP6 and the representation in Eq.~\eqref{decompBNP6}, using \pysecdec, for increasing values of the centre-of-mass energy $s_{12}$. For moderate kinematic configurations, we observe an improvement of more than one order of magnitude for most requested precision levels, from avoiding contour deformation. In the high-energy limit, this improvement is more significant, with contour deformation failing to converge beyond very low precision levels.  

\subsection{Massive triangle}
\label{sec:triangle}
In Ref.~\cite{Jones:2025jzc}, the resolutions of the massive examples relied on geometric visualisation to understand how to remap $\mathcal{F}(\mathbf{x};\mathbf{s}) = 0$ to the integration boundaries. In the massive case, this essentially limited the method to integrals with four or fewer propagators.
With this example, we will demonstrate how using GCAD can lift this limitation, by obtaining a resolution for the all-massive triangle without relying on any visualisation of the $\mathcal{F}(\mathbf{x};\mathbf{s}) = 0$ hypersurface. The integral we wish to consider is the all-massive triangle (see Figure~\ref{fig:examples})
\begin{align}
\label{eq:triangle}
J_{\mathrm{tri}}(p^2)&=-\Gamma\left(1+\epsilon\right)\lim_{\delta\to0^+}I_{\mathrm{tri}}(p^2;\delta),\\I_{\mathrm{tri}}(p^2;\delta)&=\int_{\mathbb{R}^3_{\geq0}}\!\!\mathrm{d}x_1\mathrm{d}x_2\mathrm{d}x_3\frac{\left(x_1+x_2+x_3\right)^{-1+2\epsilon}}{\left(-p^2x_1x_2+m^2\left(x_1+x_2+x_3\right)^2-i\delta\right)^{1+\epsilon}}\delta\left(1-\alpha(\mathbf{x})\right).
\end{align}
Immediately applying GCAD to reduce for the negative contribution, i.e. $\mathcal{F}_{\mathrm{tri}}(\mathbf{x};\mathbf{s}) < 0$, yields a valid resolution in terms of constraints on $x_2$ and $x_3$, although they involve algebraic functions. For example, the constraint on $x_3$ is,
\begin{equation}
    0 < x_3 < -x_1 - x_2 + \sqrt{\frac{p^2 x_1 x_2}{m^2}}.
\end{equation}
This is not ideal as the square root will appear in the resolved integrands after the upper boundary has been remapped to infinity. The issue is that in current public codes, integrands involving such square roots cannot be automatically sector decomposed. We return to this point in Section~\ref{sec:summary}, but for now we will demonstrate how, in some cases, these square roots can be avoided by a good choice of Dirac-delta function.

We make the symmetric choice by picking $\delta\left(1-x_1-x_2-x_3\right)$, which is often a good idea for one-loop integrals in general, as it results in setting $\mathcal{U}(\mathbf{x}) \to 1$. We can introduce $\beta^2=~\frac{p^2-4m^2}{p^2}\in~\left(0,1\right)$, and after integrating over $x_3$ using the $\delta$ function, we obtain the following form for the triangle integral
\begin{align}
    I_{\mathrm{tri}}=\left(\frac{1-\beta^2}{m^2}\right)^{1+\epsilon}\tilde{I}_{\mathrm{tri}}, \quad \quad \tilde{I}_{\mathrm{tri}}=\int_{\mathbb{R}^2_{\geq0}}\!\!\mathrm{d}x_1\mathrm{d}x_2\,\theta\left(1-x_1-x_2\right)\left(1-\beta^2-4 x_1 x_2-i\delta\right)^{-1-\epsilon},
\end{align}
where the appearance of $\theta\left(1-x_1-x_2\right)$ is a result of our symmetric choice of $\delta$ function. We can map the integration domain to the unit square using the variable transformation $x_2\rightarrow\left(1-x_1\right)x_2$,
\begin{equation}
\label{eq:mod_tri}
    \tilde{I}_{\mathrm{tri}}=\int_{0}^{1}\!\!\mathrm{d}x_1\mathrm{d}x_2\left(1-x_1\right)\left(1-\beta^2-4 \left(1-x_1\right)x_1x_2-i\delta\right)^{-1-\epsilon}.
\end{equation}
We now again apply GCAD, this time to reduce $\Tilde{\mathcal{F}}_{\mathrm{tri}} < 0$, where $\Tilde{\mathcal{F}}_{\mathrm{tri}}$ is the modified denominator polynomial of Eq.~\eqref{eq:mod_tri}. The result is two constraints, 
\begin{equation}
\label{eq:constraints}
    \frac{1-\beta}{2} < x_1 < \frac{1+\beta}{2} \quad \mathrm{and} \quad \frac{1-\beta^2}{4 x_1(1-x_1)} < x_2 < 1,
\end{equation}
which involves only rational functions. We start by shifting the lower boundary of $x_2$ to $0$ (while keeping the boundary at $x_2 = 1$ fixed), by making the variable transformation,
\begin{equation}
    x_2\rightarrow x_2+\left(1-x_2\right)f\left(x_1\right), \quad f\left(x_1\right) = \frac{1-\beta^2}{4 x_1(1-x_1)}.
\end{equation}
By design of the GCAD algorithm, this does not interfere with the constraint on $x_1$, and we can proceed by shifting the lower and upper boundaries of the constraint on $x_1$, to $0$ and $1$ respectively,
\begin{equation}
    x_1 \to \frac{1}{2} + \beta (x_1 - \frac{1}{2}).
\end{equation}
After applying these transformations to the integrand and including the corresponding Jacobians, the result for the negative contribution is (with the minus sign factored out already),
\begin{equation}
    \tilde{I}_{\mathrm{tri}}^{-}=2^{-1-2\epsilon}\beta^{1-2\epsilon}\int_{0}^{1}\!\!\mathrm{d}x_1\mathrm{d}x_2\left(1-x_1\right)^{-\epsilon}x_1^{-\epsilon}x_2^{-1-\epsilon}\left(1-\left(1-2x_1\right)\beta\right)^{-1}.
\end{equation}
The above procedure should then be repeated by using GCAD to reduce $\Tilde{\mathcal{F}}_{\mathrm{tri}} > 0$, which in this case yields three positive contributions, in the form of three sets of inequalities\footnote{Logically: $\Tilde{\mathcal{F}}_{\mathrm{tri}} > 0 \Rightarrow S^+ \Leftrightarrow S_1^+ \vee S_2^+ \vee S_3^+$, and for the integral: $\int_{S^+} = \int_{S_1^+} + \int_{S_2^+} + \int_{S_3^+}$.}, each in a form similar to Eq.~\eqref{eq:constraints}. The corresponding transformations are applied to each piece individually, and the final results for the three positive pieces are,
\begin{align}
    \tilde{I}_{\mathrm{tri}}^{+,1}&=\frac{1}{4}\left(1-\beta\right)^{-\epsilon}\int_{0}^{1}\!\!\mathrm{d}x_1\mathrm{d}x_2\left(2-\left(1-\beta\right)x_1\right)\left(1+\beta-\left(2-\left(1-\beta\right)x_1\right)x_1x_2\right)^{-1-\epsilon},\\
    \tilde{I}_{\mathrm{tri}}^{+,2}&=\frac{\beta}{2}\left(1-\beta^2\right)^{-\epsilon}\int_{0}^{1}\!\!\mathrm{d}x_1\mathrm{d}x_2\left(1-x_2\right)^{-1-\epsilon}\left(1-\left(1-2x_1\right)\beta\right)^{-1},\\
    \tilde{I}_{\mathrm{tri}}^{+,3}&=\frac{1}{4}\left(1-\beta\right)^{1-\epsilon}\int_{0}^{1}\!\!\mathrm{d}x_1\mathrm{d}x_2\left(1-x_1\right)\left(1+\beta-x_2\left(1-x_1\right)\left(1+\beta+\left(1-\beta\right)x_1\right)\right)^{-1-\epsilon}.
\end{align}
The total result for the resolved triangle is
\begin{align}
    J_{\mathrm{tri}}(\beta)&=-\Gamma\left(1+\epsilon\right)\lim_{\delta\to0^+}I_{\mathrm{tri}}(\beta;\delta), \nonumber \\
    I_{\mathrm{tri}}(\beta;\delta)&=\sum_{n_+=\ \!\!1}^{3}I_{\mathrm{tri}}^{+,n_+}(\beta)+\left(-1-i\delta\right)^{-1-\epsilon}I_{\mathrm{tri}}^{-}(\beta),
    \label{trisum}
\end{align}
which is the same as in Ref.~\cite{Jones:2025jzc}, but in this case obtained without relying on any visualisation of the $\mathcal{F}(\mathbf{x};\mathbf{s}) = 0$ hypersurface.

\begin{figure}[h!]
\centering
    \includegraphics[width=0.9\textwidth]{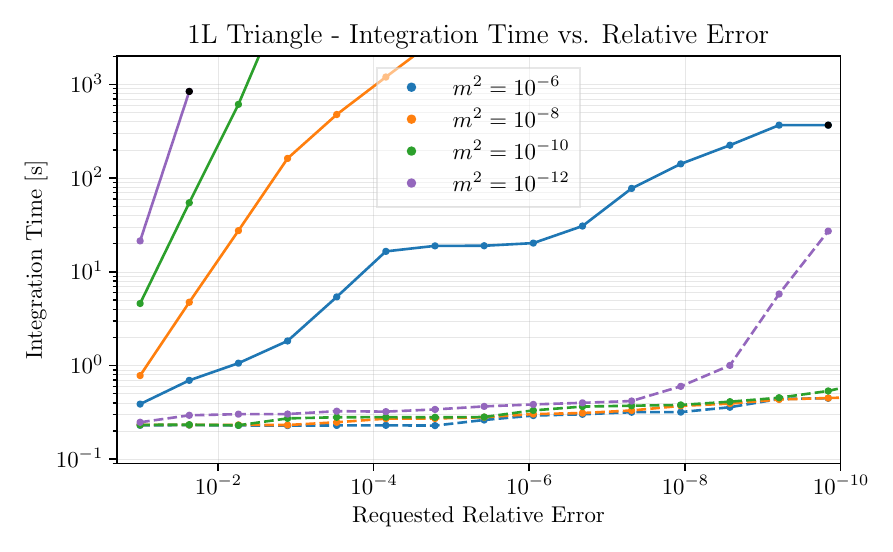}
    \caption{Timings with (solid lines) and without (dashed lines) contour deformation for the all-massive 1-loop triangle, expanded up to order $\epsilon^4$. Evaluated for different values of $m^2$ with $s_{12}=1$ fixed. The black dots on the CD lines indicate that no higher accuracy could be reached within 5 hours.}
    \label{fig:triangle1L3m_timings_masses}
\end{figure}

In Figure~\ref{fig:triangle1L3m_timings_masses}, we compare the integration time to obtain a certain requested precision, between the contour deformed triangle and the representation in Eq.~\eqref{trisum}, using \pysecdec, in the small-mass regime. In the limit $m \to 0$ the integral approaches an endpoint singularity, which pinches the integration contour and produces large cancellations. The contour deformed integral therefore performs arbitrarily badly close to this configuration, while the resolved integral is mostly immune to this problem. Indeed, in Figure~\ref{fig:triangle1L3m_timings_masses}, we observe order-of-magnitude improvements in the integration time from avoiding contour deformation in the small-mass regime, and they are most significant for the smallest $m^2$. 

\section{Summary \& Outlook}
\label{sec:summary}
We have presented a review of, and recent progress on, our method of writing Feynman integrals as a sum of integrals over real, non-negative integrands multiplied by complex prefactors. This representation eliminates the need for a contour deformation during numerical integration, which accelerates evaluation times by several orders of magnitude, in many cases. We have used the GCAD algorithm, as implemented in \texttt{Mathematica}, to reduce the systems of polynomial inequalities required for the resolutions. This allows us to resolve massive integrals without relying on a visualisation of the $\mathcal{F}(\mathbf{x};\mathbf{s}) = 0$ hypersurface, as was done in Ref.~\cite{Jones:2025jzc}. In principle, this algorithm generalises our approach to Feynman integrals with an arbitrary number of propagators. 

In practice, reducing $\mathcal{F}(\mathbf{x};\mathbf{s}) > 0$ and $\mathcal{F}(\mathbf{x};\mathbf{s}) < 0$ to constraints on $\mathbf{x}$ is very difficult for integrals with many propagators and kinematic invariants, due to GCAD scaling very poorly with respect to the number of variables in the problem. 
In the future, it will be interesting to investigate whether this can be circumvented, in particular for the special case of Feynman integrals. Furthermore, as was demonstrated in Section~\ref{sec:triangle}, the naive application of GCAD to massive integrals results in integrands containing algebraic functions, such as square roots. Since we almost always want to apply sector decomposition to the resolved integrals, these square roots are currently a problem. Extending sector decomposition in \pysecdec to handle algebraic integrands is therefore crucial to generalise the method to any Feynman integral.
Finally, the results presented here are based on timings using \pysecdec, which is optimized for contour deformed integrands. Improving the treatment of real, non-negative integrands in \pysecdec is likely to bring significant benefits in the evaluation time for our resolved integrals. 

\acknowledgments
This research was supported in part by the UK Science and Technology Facilities Council under contracts ST/X000745/1 and ST/X003167/1, in part by the Excellence Cluster ORIGINS funded by the Deutsche Forschungsgemeinschaft (DFG, German Research Foundation) under Germany’s Excellence Strategy – EXC-2094-390783311 and in part by the European Research Council (ERC) under the European Union’s research and innovation programme grant agreements 949279 (ERC Starting Grant High-PHun). SJ is additionally supported by a Royal Society University Research Fellowship (URF/R1/201268, URF/R/251034).

\bibliographystyle{JHEP}
\bibliography{main}

\end{document}

%% file: all.tikzstyles
\tikzstyle{blank}=[fill=white, shape=circle, draw=white, inner sep=0.8pt]
\tikzstyle{dot}=[fill=black, shape=circle, draw=black, inner sep=0.8pt]
\tikzstyle{fat}=[fill=blue, shape=circle, draw, line width=1pt, inner sep=0.8pt, minimum size=4mm]
\tikzstyle{blob}=[fill=gray, shape=circle, draw=black, line width=1pt, inner sep=0.8pt, minimum size=1cm]
\tikzstyle{star}=[fill=black, shape=star, inner sep=0.8pt, minimum size=5mm]
\tikzstyle{arrow}=[{-{Classical TikZ Rightarrow[length=2mm,width=1.5mm]}}, draw={rgb,255: red,61; green,171; blue,83}, line width=1pt, preaction={{draw=white,line width=2pt}}, line cap=rect]

\tikzstyle{gluoncoil}=[-, decorate, decoration={coil,aspect=1.4,segment length=2.5mm}]
\tikzstyle{fermion}=[-, draw={rgb,255: red,176; green,36; blue,39}, line width=1pt, preaction={{draw=white,line width=2pt}}, line cap=rect, postaction=decorate, decoration={markings,mark=at position .60 with {\arrow{Stealth[round,width=5pt]}}}]
\tikzstyle{scalar}=[-, line width=1pt, dashed, draw]
\tikzstyle{fermionarrow}=[-, postaction=decorate, decoration={markings,mark=at position .60 with {\arrow{Stealth[round,width=5pt]}}}]
\tikzstyle{ct scalar}=[-, line width=1pt, dashed, draw={rgb,255: red,102; green,102; blue,102}, postaction=decorate, decoration={markings,mark=at position .50 with {\node[style=star,minimum size=5mm]{};}}]
\tikzstyle{ct fermion}=[-, line width=1pt, preaction={{draw=white,line width=2pt}}, line cap=rect, postaction=decorate, decoration={markings,mark=at position 0.25 with {\arrow{Stealth[round,width=5pt]}}, mark=at position 0.75 with {\arrow{Stealth[round,width=5pt]}}, mark=at position .50 with {\node[style=star,minimum size=5mm]{};}}]